\documentclass[%
preprint,
superscriptaddress,
longbibliography,
prb,
]{revtex4-1}

\usepackage{amsmath}    
\usepackage{graphicx}   
\usepackage{verbatim}   
\usepackage{color}      
\usepackage{subfigure}  
\usepackage{hyperref}   
\usepackage{rotating}
\usepackage{multirow}

\begin{document}

\title{Seeing the real world: Comparing learning from verification labs and traditional or enhanced lecture demonstrations}
\author{Emily M. Smith}
\author{N.G. Holmes}
\affiliation{Laboratory of Atomic and Solid State Physics, Cornell University, Ithaca, NY 14853}

\date{\today}

\begin{abstract}
Instructors in introductory physics courses often use labs and demonstrations to reinforce that the physics equations introduced in lectures and textbooks describe what actually happens in the real world. The surface features and instructional learning goals of such labs and demonstrations may seem quite similar (predict-observe-explain). Nonetheless, physics education research has found very different impacts of these instructional methods on students' physics content knowledge that depend critically on several instructional characteristics. In this paper, we disentangle the research by comparing details of the learning methods in verification labs, traditional lecture demonstrations, and enhanced lecture demonstrations. We discuss possible mechanisms for the measurably different learning outcomes by dissecting the activities according to the role of prediction, cognitive load, and engagement. We use this characterization to motivate rethinking the goals of labs towards experimentation skills and beliefs, rather than for verifying physics content.  
\end{abstract}

\maketitle 

\section{Introduction}
In this paper, we discuss the similarities and differences between three instructional methods designed to reinforce physics concepts: traditional lecture demonstrations, enhanced lecture demonstrations, and verification labs. Traditional lecture demonstrations are ones where an instructor uses physical equipment to demonstrate a phenomenon from class, with little to no student participation. Enhanced lecture demonstrations, in contrast, are ones that engage students in structured forms of predicting, observing, and explaining the demonstration. In both cases, the instructor controls the equipment and the outcome is predetermined by the instructor. Verification labs similarly have predetermined outcomes, but the equipment is controlled by the students. In all three cases, the learning goal is for students to ``see'' physics in the real world. We believe that labs are an essential component of a physics education and intend for this paper to provide further evidence for why labs should be used to teach students about the nature of science and develop their experimentation and critical thinking skills, rather than to reinforce content knowledge.

Recently published research has found that verification labs do not add to students' learning of physics content in a measurable way, despite explicit goals to do so.~\cite{Holmes2017, Wieman2015} These results may seem puzzling when compared to enhanced lecture demonstrations, which share common surface features with verification labs but have been shown to improve student learning.~\cite{Crouch2004, Miller2013} In both activities, students must predict outcomes, observe the physics, and reflect on their observations with a final explanation or concept. They both involve interacting with peers and provide experiences that may be deliberately counter-intuitive, but eventually demonstrate that ``physics works.'' The stark contrast between these research outcomes (substantial learning gains in one context but not the other) raises questions: why are these similar activities yielding such different learning outcomes? In this paper, we compare verification labs to enhanced lecture demonstrations to decompose when and why these instructional methods do and do not impact learning of physics material. We also compare the characteristics of traditional lecture demonstrations as a contrast.

We have three intended aims for decomposing these instructional methods: (1) to draw attention to the underlying features of successful instructional methods; (2) to communicate the value of evaluating the foundational features of instructional methods, especially those that may appear similar; and (3) to raise awareness of the ongoing discussion among physics educators about effectively using labs to teach students skills that are unique to labs without detriment to their learning of physics content. We approach this discussion by comparing foundational components of three instructional methods to provide possible---but not comprehensive---research-based explanations for why enhanced lecture demonstrations produce measurable learning outcomes but verification labs and traditional lecture demonstrations do not. We hope this article generates a dialogue around these methods and do not intend for it to be an extensive theoretical analysis. First, we clarify our definitions of the three teaching methods.

\subsection{Definitions}
Many introductory physics laboratories, such as those discussed in the studies in Refs.~\onlinecite{Holmes2017} and~\onlinecite{Wieman2015}, use hands-on activities as a way to reinforce physics content and to demonstrate that physics applies to a real-world setting. The labs in these particular studies, like many traditional physics labs, used well-defined, structured experimental protocols that aimed for students to verify physics ideas introduced in the lecture portion of the course. The instructional materials required students to engage in structured sequences of using concepts or equations to predict the expected outcome, observe the outcome, and explain the results. These labs often prompted students to explain how the experimental data verifies physics concepts, such as by comparing whether the expected outcome aligned with an observation or explaining results that were not as expected. We refer to these types of labs as \emph{verification labs}. In this paper, we limit our definition of verification labs to those that aim to reinforce physics content previously introduced in lecture portions of the class; labs designed to have students discover physical relationships prior to instruction are not included in this definition.

\emph{Enhanced lecture demonstrations} are superficially similar to verification labs, though the equipment is handled by the instructor rather than the students. Students make a prediction about the outcome, observe the demonstration, and then may explain---to themselves or to others---the physics that explains the observation. As noted below, the role of the prediction in the case of enhanced lecture demonstrations differs from the role of generating expected results in verification labs. \emph{Interactive Lecture Demonstrations} are a specific type of enhanced lecture demonstration that include eight steps: a description of the demonstration, an individual prediction, student discussion about their predictions, a class discussion about students' predictions, a revised individual prediction, the demonstration, a class discussion and individual writing about the results, and a class discussion of analogous physical situations.~\cite{Sokoloff1997} Enhanced lecture demonstrations do not necessarily include all components of Interactive Lecture Demonstrations, such as peer discussion or the discussion of analogous physical situations. Enhanced and Interactive lecture demonstrations contrast with \emph{traditional} lecture demonstrations where students observe the demonstration with no requirement to generate or record predictions or explanations or discuss with peers. 

\begin{figure}
\includegraphics[width=\linewidth]{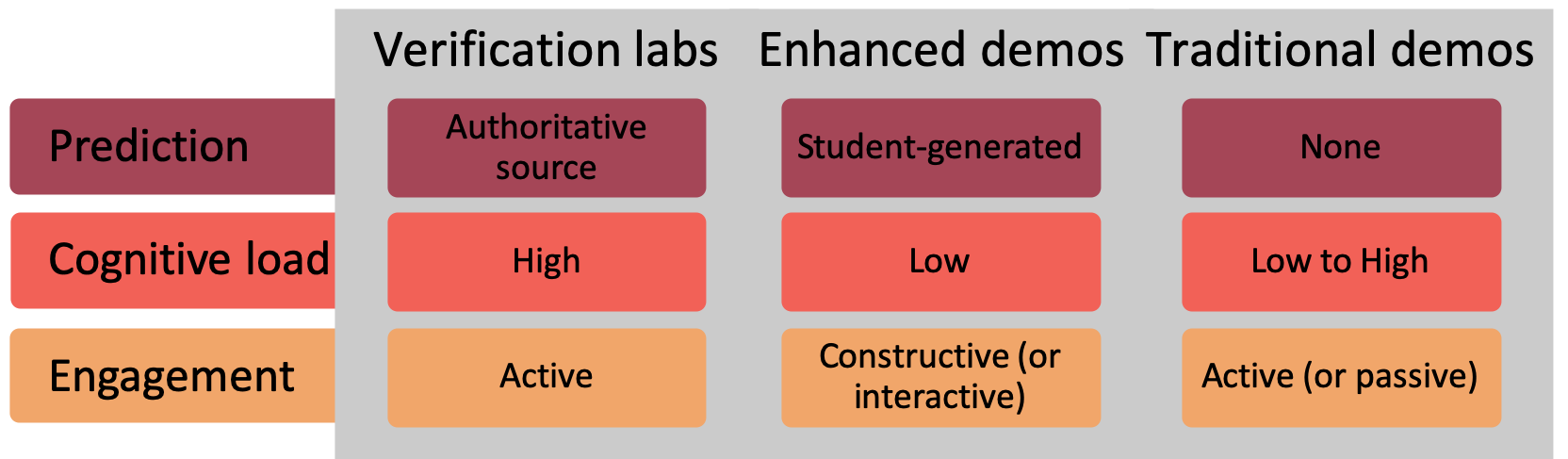}
\caption{Summary of the differences between verification labs and enhanced lecture demonstrations.}\label{fig:fig}
\end{figure}

In what follows, we evaluate \emph{verification labs}, \emph{enhanced lecture demonstrations}, and \emph{traditional lecture demonstrations} according to the role of prediction, cognitive load, and student engagement (summarized in Figure~\ref{fig:fig}). These facets provide a sufficient, albeit not comprehensive, comparison to explain the observed differences in learning from research on these methods. Dissection of an instructional method in terms of the mechanisms for learning is critical to understanding the results of its impact and for successful implementations or adaptations of that instructional method.~\cite{Borrego2013} While we discuss verification labs, we do not intend to generalize to all types of labs. Many labs aim to teach lab skills and the nature of science, rather than physics content, and so this discussion does not apply to those labs. Throughout this paper, we distinguish between traditional, enhanced, and Interactive lecture demonstrations, outlining how the differences between them relate to the measured learning outcomes. 

\section{Predictions}

Research has suggested that the generation of a prediction may be the critical component to measured gains in students' content knowledge from enhanced lecture demonstrations.~\cite{Crouch2004,Miller2013} Both verification labs and enhanced lecture demonstrations ask students to make predictions using a number of representations (mathematical, conceptual, etc.). Traditional lecture demonstrations do not include a prediction, which is a distinguishing feature from enhanced lecture demonstrations.

Enhanced lecture demonstrations require students to make an initial prediction about an overall result using their real-world experiences, their intuition, prior physics instruction---whatever resources they decide to draw on. They then must commit to their prediction by recording it, either through an open-response, closed-response, or combination-response format.~\cite{Crouch2004} 

Experiencing the demonstration provides feedback so that students must confront their recorded ideas, providing opportunities for them to make modifications, as appropriate. The activity for the demonstration is selected purposefully to address confusing and counter-intuitive concepts, often necessitating such modifications. The predictions are designed so that a student synthesizes several ideas to generate new knowledge or explores coherence between pieces of knowledge that were unconnected. Engaging students with a prediction before observing and explaining the demonstration also serves to initiate or activate an organizational structure for their knowledge,~\cite{Chi1981} creating a so-called ``Time for Telling.''~\cite{Schwartz1998} In Interactive Lecture Demonstrations, a critical step after discussing the results of the demonstration is to discuss analogous physical situations so that students can further build up and generalize their knowledge organization.~\cite{Sokoloff1997}

Verification labs use predictions in a different way: Students most often apply previously learned knowledge and equations to find an expected value or result. Students are rarely, if ever, expected to contribute their individual knowledge or ideas in generating these expectations. Students are often guided to use particular representations that are deemed appropriate by the instructor or the written instructions, often equations or graphs using specified parameters. 

One important distinction between the use of predictions in these ways is the source of the prediction. The purpose of a verification lab experiment is to verify that the previously introduced physics is ``true'' in the real world. The predicted outcomes, therefore, come from applying authoritative sources---lecture, teaching assistants, or the textbook---rather than being generated from students' individual ideas. Within the verification lab experience, students may never directly confront their own incorrect or novice-like ideas, unlike in enhanced lecture demonstrations.

Another important consideration may be the timing of the prediction. In verification labs, the predictions are generally made after instruction on the topic, so that predictions are applications of previously learned ideas. At times, course timing may necessitate verification labs to precede instruction, but the intention is that the students verify concepts seen previously in class (by definition). A demonstration, however, may be shown to students before or after instruction on the topic. It is therefore unclear whether the timing of the activity (or the timing of the prediction) is important for understanding the impacts on student learning. 

\section{Cognitive load}

Cognitive load refers to the notion that an individual has a capacity (i.e., limit) for processing information in their short-term memory.~\cite{Sweller1988} It is supported by research suggesting that individuals can only store so much in their memory at once or can only pay attention to so many things at a time. The way in which information is presented to learners, the way learners interact with the information, or the sheer amount of information presented to learners can impact their cognitive load.~\cite{Paas2003} Simplistically, high cognitive load impedes learning.~\cite{Kirschner2006} 

In any real-world, hands-on activity---be it a lab or a demonstration---many features contribute to a student's cognitive load. In a verification lab, students must coordinate equipment, physics theory, equations, concepts, data and its variability, a notebook or worksheet, and the complexities of team work, among other tasks. In enhanced lecture demonstrations, much of that work is done for the students. Their main cognitive task is making predictions and coordinating their predictions with their observations. Enhanced lecture demonstrations also deliberately structure students' cognitive load using a series of instructional steps that narrow students' focus to specific tasks: making a prediction, comparing predictions with observations, and explaining the physics of the observed demonstration. The deliberate sequencing of instructional decisions used in enhanced and Interactive lecture demonstrations serve to focus student attention toward learning the physics at hand. Interactive Lecture Demonstrations also intentionally hide distracting components of the equipment, so that student attention is focused towards the intended learning objectives, removing unnecessary cognitive load.~\cite{Thornton1997}

We argue that a verification lab is a high cognitive load activity, while an enhanced lecture demonstration is a much lower cognitive load activity. Traditional lecture demonstrations may be high or low cognitive load, depending on the demonstration and associated equipment.

Some education researchers have argued that high cognitive load does not necessarily impede student learning.~\cite{Kapur2016,Schwartz2011} In those studies, ill-structured problems with arguably high cognitive load were found to improve student learning over highly structured activities with lower cognitive load. The effects of cognitive load were found to depend on how the additional load relates to the knowledge being acquired: \emph{germane or effective} cognitive load is related to the content while \emph{extraneous or ineffective} cognitive load is unrelated.~\cite{Paas2003} Learning may be enhanced by low extraneous cognitive load or high germane cognitive load. Much of the cognitive load in verification labs and traditional lecture demonstrations is extraneous and verification labs typically employ high levels of structure to reduce the extraneous cognitive load. 

\section{Students' cognitive engagement}

Students can be engaged during class in a number of different ways depending on the type of activity. Chi and colleagues have hierarchically ordered four types of activities based on the cognitive engagement the activities elicit: 

\begin{center} \emph{interactive} $>$ \emph{constructive} $>$ \emph{active} $>$ \emph{passive}.~\cite{Chi2009,Chi2014,Chi2018} \end{center}

\noindent All types of activities assume that students are on-task and that their behaviors are oriented to engaging with the instructional materials.~\cite{Endnote} In \emph{passive activities} students are passive and unlikely to connect new information with their existing knowledge. In \emph{active activities} students participate through physical actions such as manipulating objects, gesturing, or highlighting, but without generating new knowledge. The physical actions are expected to activate students' existing knowledge, but not to provide new information beyond what is provided to students from the instructional materials. In \emph{constructive activities}, students generate new knowledge through behaviors such as explaining a concept, generating a new prediction that connects or synthesizes knowledge to produce knowledge that is new to the student (not to apply previous knowledge directly), or making new connections. In \emph{interactive activities}, students use a partner's contributions to build new knowledge, through behaviors such as arguing and revising from feedback. By definition, each category contains the previous categories, so interaction requires both partners to engage in construction. Interactive activities are thought to produce greater learning gains than constructive activities because they require students to use shared ideas to construct new knowledge from multiple sources and perspectives.~\cite{Chi2014,Chi2018} 

For example, direct instruction (i.e. lecturing) may be categorized as: (1) a \emph{passive} activity if, on average, students are listening but not taking notes, (2) an \emph{active} activity if students are copying notes from a board, or (3) a \emph{constructive} activity if students are writing notes that self-explain the presented material.~\cite{Chi2018} In this framework, most research-based instructional strategies in physics education may be categorized as \emph{constructive} or \emph{interactive} activities. 

Because students may observe a demonstration and copy notes about the physics that is presented, we conservatively categorize traditional lecture demonstrations as active activities. In practice, students may view traditional lecture demonstrations merely as entertainment, a passive activity.~\cite{Miller2_2013} 

Enhanced lecture demonstrations are constructive because students receive feedback and/or new knowledge from their peers and the demonstration and, as a result, are able to revise their understanding of the situation. Interactive Lecture Demonstrations incorporate additional interactive elements such as redeveloping a prediction following peers' contributions. Often, students discuss again with their peers after observing the demonstration to generate an explanation before receiving instruction. Interestingly, the measured learning differences between enhanced and Interactive lecture demonstrations are statistically small or, perhaps, insignificant.~\cite{Crouch2004} Evaluations of students' discussions during clicker questions have shown that not all students contribute reasoning to the group discussion,~\cite{Knight2013} which may explain these differences. That is, peer discussions may be constructive, but not necessarily interactive activities. 

As defined, verification labs certainly meet the criteria for active activities. Students use their hands to work with equipment, answer questions, and often manipulate data and equations. Generally, verification labs are also meant to be constructive or interactive; students work in pairs or groups (suggesting interactivity) and have opportunities to construct explanations of the data according to relevant physics concepts. Intent and opportunities for construction and interactivity, however, do not necessarily translate into action (and the measured differences between traditional and enhanced lecture demonstrations is a clear example of this). In verification labs, the connections between the experimental apparatus and mathematical formulation are typically detailed in the instructions. Predictions, reflection questions, or explanations more often require application of previously learned concepts or procedures, rather than generation of new ones. Without construction of new knowledge, there is no interactivity (again, interactive does not just mean talking to someone else in this context). 

\section{Discussion}

In this paper, we have outlined possible mechanisms for the differences in learning outcomes between verification labs, traditional lecture demonstrations, and enhanced lecture demonstrations. In verification labs, students base their predictions on authoritative sources, have many experimental features they must attend to, and do not typically engage in constructive or interactive learning. In traditional lecture demonstrations, students do not make predictions, have varied cognitive load depending on the activity, and engage actively or passively. In enhanced lecture demonstrations students, generate an individual prediction, are able to focus attention on key aspects, and engage in constructive or interactive learning. 

There exist many other possible variables to consider and limitations to the arguments above. For example, the role of the prediction in enhanced lecture demonstrations is to induce conceptual change:~\cite{Strike1982} Students confront their incorrect ideas (misconceptions) recorded in their prediction and change their ideas as a result of new, convincing evidence. But research repeatedly demonstrates that the process of conceptual change can be complicated by students' motivation, values, and beliefs.~\cite{Pintrich1993} 

We have also mentioned that there is conflict in the literature about whether and when high or low cognitive load is beneficial to students' learning. We are unaware, however, of any studies that have measured students' cognitive load during physics labs or demonstrations. Two studies aimed to measure students' cognitive load in chemistry labs,~\cite{Hubacz2004,Winberg2007} but found many confounding factors, highlighting the difficulties of measuring cognitive load in classroom environments.

Few studies have examined students' modes of cognitive engagement in labs or lecture demonstrations. One study hints that students' cognitive engagement in labs may drastically vary from high engagement to low engagement to enthusiastic participation in the activity without cognitive engagement.~\cite{Schmidt2017} However, we are uncertain how these levels of cognitive engagement translate to different populations and different types of labs or lecture activities. 

We have also not discussed the role of the instructor in these settings. Verification labs are often facilitated by a teaching assistant, while enhanced lecture demonstrations are typically facilitated by a faculty instructor. Teaching assistants in verification labs have little autonomy to make instructional decisions regardless of their training as instructors. Students are primarily interacting with the written instructional materials in the highly guided verification labs discussed here and it is unclear how to distinguish the impact of the teaching assistant from the impact of the instructor who originally wrote the lab materials.

For better or for worse, assessment also influences what students do, how they act, and to what they pay attention.~\cite{Shinske} In lecture demonstrations and labs, different assessment strategies may encourage students to focus on different aspects of the activities. However, there are countless variations in how students may be assessed in both instructional methods. Typically, student participation from enhanced lecture demonstrations is required for the activity to progress due to cycles of self-revision. Whereas in verification labs, student participation may not be built on or discussed because verification labs give students an optimal procedure to follow; their participation may be motivated by correctly carrying out (performing) the procedure, rather than mastering the content.~\cite{Dweck1986} In enhanced lecture demonstrations, however, the focus is usually on developing ideas, perhaps in a mastery, growth-mindset way:~\cite{Dweck1999} It's alright if we make a wrong prediction, as long as we can make sense of it afterwards. Traditional lecture demonstrations are typically not assessed.

We hope researchers will formally test these ideas. How might verification labs impact learning if students do not yet know the concept they are attempting to verify and can meaningfully construct the knowledge (and predictions)? How do different assessment structures impact learning? Are there ways to reduce unnecessary cognitive load in labs to improve learning? What structural manipulations would improve interactivity? 

We are aware that physics education researchers may view the comparison that we presented as obvious and unnecessary. It is important, however, to compare different approaches to instruction that share foundational and/or superficial similarities so that we can dive into the mechanisms for students' success in physics. We view this paper as a start to that conversation, using an example of superficially similar but foundationally different instructional strategies that are relevant and accessible to physics education researchers and research-informed instructors alike.

\subsection{Implications for instruction}

\begin{figure*}
\includegraphics[width=0.5\textwidth]{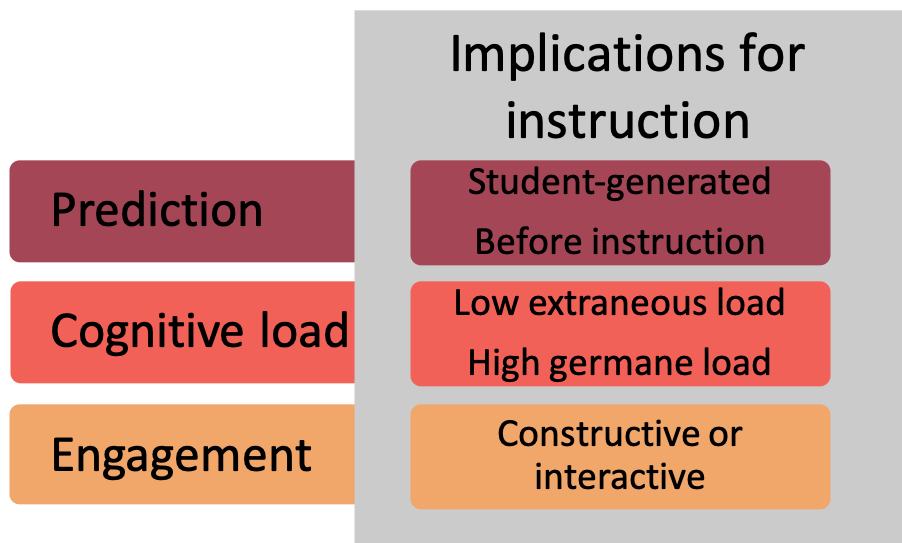} \caption{Implications for instruction suggested from the arguments in the paper.}\label{fig:fig2}
\end{figure*}

From this analysis, we could extract general implications for instruction, be it labs or demonstrations (Figure~\ref{fig:fig2}). While these suggestions have already been shown to be effective in the case of enhanced and Interactive lecture demonstrations, verification labs have inherent limitations. Scheduling labs so that students have not yet received instruction, but soon will, is complicated with large multi-section courses. Training lab instructors to facilitate constructive and interactive engagement is non-trivial.~\cite{Chase2019} Removing the activities associated with the extraneous cognitive load may remove much of the valuable aspects of experimentation.

A key question concerns resources and efficiency: are such modifications to verification labs, while maintaining the verification goals, worth the effort? The literature suggests that learning could be more efficiently achieved through a 15-20 minute enhanced (or Interactive) lecture demonstration, rather than a two- to three-hour lab. Our discussion of cognitive load highlighted that many aspects of labs do not exist in demonstrations or other forms of instruction, such as coordinating equipment, analyzing data and managing its variability, navigating team work, and communicating the results of an experiment. These features are valuable instructional goals in and of themselves.~\cite{Hofstein2004}

As argued elsewhere,~\cite{NGSS,JTUPP,AAPTLabs} labs may be more effectively used for focusing on skills that can be uniquely learned through the activity associated with labs. The American Association of Physics Teachers has compiled a set of recommendations for undergraduate physics labs that provide a starting point for instructors.~\cite{AAPTLabs} The Physics Teacher and the American Journal of Physics also have many articles with practical applications of those ideas, such as those in Refs.~\onlinecite{HolmesPendulum,Etkina2002,Buffler2008,Kung2005,Moore2014, HolmesLearningGoals}.

\begin{acknowledgments}
This work is partially supported by the Cornell University College of Arts and Sciences Active Learning Initiative. We would also like to thank the feedback and contributions of the Cornell Physics Education Research Lab, Dr. Peter Lepage, Dr. Carl Wieman, and Dr. Kate Follette. 
\end{acknowledgments}


\begin{thebibliography}{5}
\bibitem{Endnote}These terms are not synonymous with ambiguous terms for research-based instructional methods (e.g., active learning, interactive engagement).
\end{thebibliography}
\end{document}